\documentclass[aps,twocolumn,showpacs,preprintnumbers, floats]{revtex4}
\usepackage{amsfonts}
\usepackage{amsmath}
\usepackage{graphicx}
\usepackage{dcolumn}
\usepackage{bm}

\begin{document}

\preprint{}
\title{Universality of the Scaling Law for Ferroic Domains}
\author{G. Catalan$^{1}$}
\email{gcat05@esc.cam.ac.uk}
\author{J. F. Scott$^{1}$, A. Schilling$^{2}$}
\author{J. M. Gregg$^{2}$}
\affiliation{$^{1}$ Department of Earth Sciences, University of
Cambridge, Downing Street, Cambridge CB2 3EQ, United Kingdom}
\affiliation{$^{2}$ International Research Center for Experimental
Physics (IRCEP), Queen's University of Belfast, Belfast BT7 1NN,
United Kingdom}
\date{\today}

\begin{abstract}
We show how the periodicity of 180$^{\circ}$ domains as a function
of crystal thickness scales with the thickness of the domain walls
both for ferroelectric and for ferromagnetic materials. We derive an
analytical expression for the universal scaling factor and use this
to calculate domain wall thickness and gradient coefficients
(exchange constants) in some ferroic materials. We then use these to
discuss some of the wider implications for the physics of
ferroelectric nano-devices and periodically poled photonic crystals.

\end{abstract}

 \maketitle

The generic term "ferroic" designates crystalline materials that are
ordered either ferroelectrically, ferromagnetically or
ferroelastically (including also antiferroic configurations).
Ferroic materials usually display domains, that is, regions that are
either ordered along different polar directions or along the same
direction but with opposite polarity (180$^{\circ}$ domains). Kittel
showed some 60 years ago that the width of 180$^{\circ}$ magnetic
domains ($\textit{w}$) is correlated to the thickness of a crystal
in a very well defined manner: the square of the domain width
(\textit{w}) is directly proportional to the thickness of the
crystal (\textit{d}) \cite{Kittel46}. Kittel's law was latter
extended by Mitsui and Furuichi (1953) for ferroelectric materials
\cite{Mitsui53}, and by Roytburd (1972) for epitaxially clamped
ferroelastic ones \cite{Roytburd72}.

Recently, Schilling et al. \cite{Schilling06} have shown that the
constant of proportionality between w$^{2}$ and \textit{d} is a
defining characteristic of the type of ferroic transition being
considered, with ferromagnets having generally bigger domains than
ferroelectrics for crystals of the same thickness. These
experimental results were rationalised by Scott \cite{Scott06}, who
observed that the differences between ferroelectric and
ferromagnetic domain periodicity essentially disappeared once the
domain wall thickness was incorporated as a scaling factor.
Mathematically this was expressed as $\frac{w^{2}}{Td}=M$, where $T$
is the thickness of the domain wall and \textit{d} is the crystal
thickness; since domain walls tend to be narrow (few unit cells) for
for all ferroelectrics, and broader for ferromagnets (tens of
nanometers), the dimensionless factor \textit{M} ends up being
pretty much the same for both. This is nicely illustrated in Figure
1: the different characteristics for w$^{2}$ as a function of
crystal thickness of ferroelectrics and ferromagnets fall into the
same parent curve once the square of the domain width is scaled by
the domain wall thickness \emph{T}.

\begin{figure}[tbp]

\includegraphics[bb=0 0 400 300, width=0.85\textwidth]{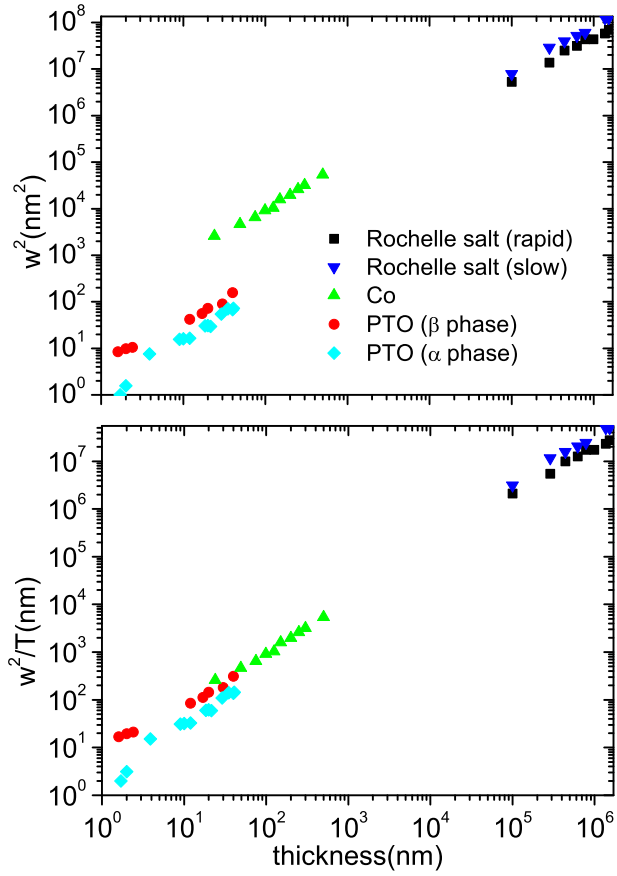}
\caption{(Color online)Above:Square of the 180$^{\circ}$ domain
width as a function of crystal thickness for some ferroics (data
extracted from refs. \cite{Mitsui53, Hehn96, Streiffer02}). Below:
When the square of the domain width is divided by the wall
thickness, all datapoints fall essentially on the same parent curve.
The domain wall size used for the scaling has been extracted from
refs. \cite{Mitsui53, Hubert98, Meyer02}}
\end{figure}

In this Letter we discuss the nature of the dimensionless constant
\textit{M}, and derive a simple analytical expression for it. We
then use the result to calculate the thickness of domain walls in
several ferroic materials, and discuss some of the wider
implications for practical applications.

We begin by writing the Landau thermodynamic potential describing a
second-order phase transition in a ferroic. For simplicity's sake,
we shall focus on the case of $180^{\circ}$ walls in a uniaxial
ferroic, so a single order parameter will suffice. We call this
order parameter \textit{Q}, and it can be either the polarization or
the magnetization, depending on the type of ferroic being
considered. Across a domain wall there is a change of the sign of
the order parameter. That means that there is necessarily a
gradient, whose associated energy must be incorporated into the
thermodynamic potential. Assuming that \textit{Q} points along the z
direction and that the domain wall is perpendicular to the
\textit{x} axis, the thermodynamic potential is
\begin {equation}
\Delta
G=\frac{a}{2}Q^{2}+\frac{b}{4}Q^{4}+\frac{k}{2}\left(\frac{\partial
Q}{\partial x}\right)^{2} \label{Landau}
\end {equation}
For a mono-domain state, and also for the center of the domains, we
can neglect the gradient term, and minimization then leads to the
familiar result for the order parameter in the ferroic state:
\begin{equation}
Q_{0}^{2}=-\frac{a}{b}
\label{orderparameter}
\end{equation}
The second derivative of the free energy with respect to the order
parameter is the stiffness. In the ferroic state, the result is
$\chi^{-1}=-2a$ or, relating this to the order parameter,
\begin{equation}
\chi_{c}^{-1}=2bQ_{0}^{2}
\label{permittivity}
\end{equation}

Here the term "stiffness" has different meanings depending on the
ferroic context, being inverse permittivity for ferroelectrics and
inverse susceptibility for magnets.

The energy density of the domain wall is calculated by minimizing
the energy difference between a mono-domain state and a state with
one domain wall. That is, one has to minimize the following
\cite{Mitsui53, Zhirnov59}:
\begin {equation}
\sigma=\int\limits_{-\infty}^{+\infty}\left(\frac{a}{2}(Q^{2}-Q_{0}^{2})+\frac{b}{4}(Q^{4}-Q_{0}^4)+\frac{k}{2}\left(\frac{\partial
Q}{\partial x}\right)^{2}\right)dx \label{LandauWall}
\end {equation}
Where $Q_{0}$ is given by eq. (2). Minimization of $\sigma$ with the
boundary condition Q(x=$\pm\infty)=Q_{0}$ leads to the solution of
the profile of the order parameter across the domain wall
Q(x)=$Q_{0}\tanh(x/\delta)$, where the characteristic thickness is
\begin{equation}
\delta=\frac{1}{Q_{0}}\sqrt{\frac{2k}{b}}=\sqrt{-\frac{2k}{a}}
\label{delta}
\end{equation}

and the energy density of the domain wall is
\begin{equation}
\sigma=\frac{4}{3}Q_{0}^{3}\sqrt{2kb}=\frac{4}{3}Q_{0}^{2}\sqrt{-2ka}
\label{wall}
\end{equation}

Several important simplifications have been made in the above
treatment. First, in constraining the order parameter to just one
dimension we have implicitly discarded the possibility of Bloch
walls and Neel walls. This is in principle wrong for magnetic
materials (though not for ferroelectrics). However, the analysis of
such walls in magnetic materials actually arrives at the same
solutions \cite{Hubert98}. Second, in limiting our Landau expansion
to the $Q^{4}$ order, we are limiting ourselves to second order
transitions, which is not the case for several important
ferroelectrics such as PbTiO$_{3}$ and BaTiO$_{3}$, although in thin
films of these two materials epitaxial clamping (regardless of
strain) changes the transition from first to second order
\cite{Pertsev98, Catalan04}. For the exact solution of the first
order domain wall the reader is adviced to look at ref.
\cite{Cross91}. And third, we have neglected the elastic coupling to
the lattice distortions (strain terms), which is particularly
important in ferroelectrics as they generally are also ferroelastic.
However, the effect of strain can be incorporated by a
renormalization of the coefficients in the Landau expansion
\cite{Zhirnov59, Lines&Glass, Pertsev98, Catalan04}, so our
treatment is still be valid once the renormalized coefficients are
used.

Regarding the physical interpretation of eq.\ref{delta}, \textit{k}
represents an "exchange" constant, as its energy contribution is
proportional to the mismatch of spins/dipoles with respect to their
neighbors (gradient term) whereas \textit{a} and \textit{b}
represent the "anisotropy" contributions, as they indicate the
strength of the alignment of the order parameter with respect to the
crystallographic axes. Quite naturally, it follows that if the
anisotropy terms are big, the domain walls will tend to be narrow so
as to minimize the number of misaligned spins/dipoles, whereas if
the exchange \textit{k} is big the domain walls will tend to be wide
so that the gradient is as small as possible. In magnets, the
exchange interaction wins, whereas in ferroelectrics the opposite is
true, hence the difference in domain wall thickness generally
observed between the two types of ferroic \cite{Zhirnov59,
Schilling06}.

The exchange constant \textit{k} is well characterized for most
magnetic materials, but that is not the case for
ferroelectrics\cite{Cao94}, a problem which has so far complicated
analysis based on eqs. \ref{domainwidth} and \ref{wall}. It is
therefore interesting to write the energy density as a function of
domain wall thickness, which removes the dependence on \textit{k}:
\begin{equation}
\sigma=\frac{4}{3}Q_{0}^{4}b\delta=-\frac{4}{3}Q_{0}^{2}a\delta
\label{sigmadelta}
\end {equation}

This expression will be used later.

We turn now to the relation between domain periodicity and thickness
in a crystal slab cut perpendicular to the polar direction. The
uncompensated dipoles/spins at the surface generate a large
electrostatic/magnetostatic energy, which is reduced by creating
domains of opposite polarity. The depolarization/demagnetization
energy of the two crystal surfaces as a function of domain width is:
\begin{equation}
F_{surface}=\frac{7\zeta(3)Q_{0}^{2}}{\pi^{3}}\sqrt{\chi_{a}\chi_{c}}w
\label{depolarization}
\end {equation}

Where $\zeta(3)$ is Riemanns zeta function
$\zeta(3)\simeq1.202$\cite{prefactor}. Although the physical forces
involved (electrostatic, magnetostatic) are different, the Maxwell
equations for the energy are analogous, and thus the resulting
expression for the surface energy  ends up being much the same
\cite{Kittel46, Mitsui53, Bjorkstam67, Kopal97, Hubert98}; the
difference between the two ferroics is thus not contained in the
shape of the equation, but only in the magnitudes involved: the
order parameter \textit{Q} (polarization/magnetization) and the
permittivity/susceptibility $\chi$.

The reduction in surface energy achieved by introducing domains is
of course partly offset by the energy cost of the domain walls,
which is proportional to $\sigma$, to the wall size (itself
proportional to the crystal thickness \textit{d}) and to the number
density of domain walls (inversely proportional to the domain width
\textit{w}). Hence, $F_{wall}=\sigma d/w$. Adding the two energy
components and minimizing with respect to the domain width
\textit{w} leads to the standard result
\begin{equation}
w^{2}=\frac{\pi^{3}\sigma\sqrt{\chi_{a}\chi_{c}}}{7\zeta(3)Q_{0}^{2}}d
\label{domainwidth}
\end {equation}

If we now substitute the order parameter $Q_{0}$ and the energy
density $\sigma$ by their respective expressions from eqs.
\ref{orderparameter} and \ref{sigmadelta}, the final expression for
the dimensionless factor is:

\begin{equation}
M\equiv\frac{w^{2}}{d\delta}=\frac{2}{3}\frac{\pi^{3}}{7\zeta(3)}\sqrt{\frac{\chi_{a}}{\chi_{c}}}\simeq2.455\sqrt{\frac{\chi_{a}}{\chi_{c}}}
\label{universal}
\end {equation}

The experimental observation that \textit{M} is generally a number
in the range 1-10 for any ferroic is thus explained: the result is
always a numerical constant ($\simeq$2.455) modified by the square
root of the susceptibility anisotropy. The appeal of this expression
is not just in its simplicity and generality, but also in that it
can actually be exploited for practical purposes.

It has been a long standing and challenging problem to establish the
thickness of domain walls in ferroics, and very specially in
ferroelectrics, as the later tend to be very thin and hard to
measure experimentally \cite{Floquet, Pam, Shilo}. On the other
hand, theoretical approaches based on phenomenological models
suffered from the fact that the coefficient \textit{k} of the
gradient term is hard to characterize experimentally. Equation
\ref{universal} dispenses the need to know such a coefficient.
Measuring domain width and crystal thickness and knowing the
dielectric constants of a material should be enough to estimate the
domain wall width. So does it work?

In figure 1 we have shown the square of the domain width as a
function of crystal thickness for 180$^{\circ}$ domains in
ferroelectric $PbTiO_{3}$ and Rochelle Salt, and ferromagnetic Co.
All of them can be analyzed with the present treatment, although in
the case of the PTO films a correction due to the effect of the
substrate must be taken into account \cite{Bjorkstam67,
Streiffer02}. The measured slope, the permittivities and the
calculated thickness of the $180^{\circ}$ domain walls are shown in
table 1, next to previous values extracted from the literature.

Our predicted value for $\delta$ of the Rochelle Salt is ~13{\AA}
(and thus the wall thickness is $T=2\delta$=26{\AA}), compatible
with the results of Mitsui and Furuichi
(T=24-47{\AA})\cite{Mitsui53}, and Zhirnov
($\delta$=12-220{\AA})\cite{Zhirnov59}. As for the predicted value
for the domain walls of PTO, once the effect of the STO substrate
has been taken into account \cite{Bjorkstam67, Streiffer02} we
obtain that $\delta$=2.45{\AA} or T=4.9{\AA}, in excellent agreement
with the first principle calculations of Meyer and
Vanderbilt\cite{Meyer02}. The combination of our model with
experimental data thus agrees with previous estimates, and support
the view that ferroelectric $180^{\circ}$ domain walls are
atomically sharp \cite{Kinase57, Lawless68}.

The above equations apply to magnetic materials too. Fitting the
data of Co to our model, the calculated domain wall thickness is
$~20nm$, which is somewhat thicker than previous theoretical
estimates that yield a value of ~5-10nm, but thinner than the
experimentally determined values of ~46nm \cite{Donnet95}. All in
all, the results suggest that the method is quite robust for the
analysis of different ferroics.

Moreover, once the domain wall thickness has been determined, one
can go back to eq. \ref{delta} and determine the value of the
exchange constant (\textit{k}) for the material, which is also a
long standing challenge in ferroelectrics \cite{Cao94}. It is worth
emphasizing here that this constant is of great importance in
determining the performance of ferroelectric thin films and
nanostructures \cite{Scott06, Dawber05}, where the gradient term
associated with the surface depolarization has a strong effect on
the functional properties \cite{Binder81, Vendik02, Zembilgotov03,
Glinchuk04, Bratkovsky06}. Our calculated values of \textit{k} for
PbTiO$_{3}$ and Rochelle salt are included in table 1.

\begin{table}
\noindent
\def\arraystretch{1.2}
\begin{tabular}{lcccccc}
     material & $\delta M$(nm) & $\epsilon_{x}$ & $\epsilon_{z}$ & $\delta(\AA)$ & $\delta_{previous}(\AA)$&\textit{k}(m$^{3}$/F)\\ \hline
  Rochelle Salt &   21  &   445   &   9.8   &   13 &12-22\textrm{\cite{Mitsui53}} &9$\times10^{-11}$\\
  $PbTiO_{3}$   &   3.5   &   124   &   66 & 2.45&$\simeq2$\textrm{\cite{Meyer02}} &2.8$\times10^{-11}\textrm{\cite{gradient}}$\\
 \\
\end{tabular}
\caption{Experimental slope of w$^{2}$ vs d for $180^{\circ}$
domains in two ferroelectrics (slope=$\delta M$), calculated
thickness of the domain walls and calculated value of the exchange
constant \textit{k}. The Landau coefficients used in the
calculations have been extracted from refs. \cite{Mitsui53} and
\cite{Pertsev98}. The domain wall thickness is compared with
previous published estimates}
\end{table}

Parenthetically, we note also that regular domains in ferroelectric
crystals have important applications in photonics, where they are
used for frequency conversion through quasi phase matching
\cite{Armstrong62}. Presently, the regularly spaced stripe domains
are achieved through periodic poling, which has limitations due to
the large coercive field -and some times finite conductivity- of
some of the most important photonic crystals, such as $LiNbO_{3}$
and $KTiOPO_{4}$ (KTP). Importantly also, artificially fabricated
domain structures are generally not in thermodynamic equilibrium,
and switchback can occur \cite{Batchko99}. In theory it should be
possible to achieve self-patterned and stable regular domains in
ferroelectric photonic crystals by cutting them at the right
thickness and preventing charge screening upon cooling through the
phase transition, although in practice periodic poling is always
likely to be required (e.g. to maximize registration). As a
practical example of reverse engineering, the measured value of the
domain wall thickness for KTP is 3\AA \cite{Pam}, while the
longitudinal and transverse relative permittivities are ~15 and ~11
respectively \cite{Bierlein96}; accordingly, a periodic domain
structure with a domain width of e.g. 5 microns would be most stable
for a crystal $\simeq$3cm thick or, conversely, a 0.5mm crystal of
KTP can have domains as small as 0.7$\mu$m. This suggests that the
known difficulty in stabilizing narrow domains in thick crystals is
not due to intrinsic factors; indeed, experiments that make use of
the depolarizing field have achieved self-patterned sub-micron
domains in LiNbO$_{3}$ \cite{Shur00}.

In sum, the universal scaling law for ferroic domains provides a
versatile and powerful tool for analyzing the physical properties of
ferroic materials in general, and ferroelectrics in particular. Our
own analysis of existing data suggests that the thickness of
$180^{\circ}$ domain walls in ferroelectrics is extremely narrow (of
the order of one unit cell), and that regular patterns of sub-micron
domains can be achieved in photonic crystals. More studies of domain
periodicity as a function of crystal thickness should be carried out
to establish domain wall thickness and exchange parameters for other
relevant ferroelectrics such as $LiNbO_{3}$ and $BaTiO_{3}$.

We thank Professor Pam Thomas for useful feedback about periodically
poled ferroelectrics. GC acknowledges financial support from the EU
under the Marie Curie Fellowship programme.

\end{document}